\documentclass[preprint,preprintnumbers, prd, floatfix,  superscriptaddress,nofootinbib] {revtex4-1}
\usepackage{graphicx}
\usepackage{epsfig}
\usepackage{bm}
\usepackage{amssymb}
\usepackage{float}
\usepackage{amsmath}
\usepackage{dcolumn}
\usepackage{cancel}
\usepackage[colorlinks]{hyperref}
\usepackage[usenames,dvipsnames]{color}
\usepackage{color}
\usepackage{epstopdf}
\usepackage{nccbbb}
\usepackage{ulem}
\usepackage{makecell}

\hypersetup{
     breaklinks=true,
    pdfstartview={FitH},    colorlinks=true,       
    linkcolor=blue,          
    citecolor=red,        
    filecolor=magenta,      
    urlcolor=blue,           
    anchorcolor=green,      
    linktocpage=true
}

\def\doi{http://doi.org}

\begin{document}

\title{Reducing the $H_0$ Tension with Exponential Acoustic Dark Energy}

\author{Lu Yin}
\email{yinlu@sogang.ac.kr}
\affiliation{Center for Quantum Spacetime, Sogang University, Seoul 04107, Korea}
\affiliation{Department of Physics, Sogang University, Seoul 04107, Korea}

\begin{abstract}
The Hubble tension arises from different observations between the late-time and early Universe.
We explore a new model with dark fluid, called the exponential Acoustic Dark Energy (eADE) model, to relieve the Hubble tension.  
The eADE model gives an exponential form of the equation of state (EoS) in the acoustic dark energy, which is the first time to explore an exponential form for the EoS. 
In this model, the gravitational effects from the acoustic oscillations of the model can impact the CMB phenomena at the matter radiation equally epoch.
We give the constraints of the eADE model by the current cosmological dataset. The comparison of the phenomena with the standard model can be shown through CMB and matter power spectra. 
The fitting results of our model have $H_0 = 70.06^{+1.13}_{-1.09}$ in 95$\%$ C.L. and a smaller best-fit value than $\Lambda$CDM.

\end{abstract}

\maketitle

\section{Introduction}
After the accelerated expanding Universe discovered in 1998\cite{Riess:1998cb, Perlmutter:1998np}, the $\Lambda$CDM has been one of the most important models to explain modern cosmology, and that can fit very well with many observations~\cite{Copeland:2006wr}.
When considering the question ``how fast is this expansion nowadays?" we need to focus on the Hubble parameter.
Unfortunately,  the debate about the value of the present Hubble constant ($H_0$) attracted much attention with the different precise observations in recent years. 
The standard candle measurements in low redshifts, such as type Ia Supernovae, get an $H_0$ equal to $74.03 \pm 1.42$ [km/s/Mpc]~\cite{Riess:2019cxk}, while the result from high redshift give an $H_0$ of $67.4 \pm 0.5$ [km/s/Mpc] in Cosmic Microwave Background's paper~\cite{Aghanim:2018eyx}.
This phenomenon is called the ``Hubble tension", which has beyond 4.4$\sigma$. 
The precision of these cosmological measurements is about 1$\%$, 
which indicates the discrepancy can not be systematic errors. 
Moreover, a model-independent technique with strong gravitational lensing experiment -- H0LiCOW -- gives $H_0 = 73.3_{-1.8}^{+1.7}$ [km/s/Mpc]~\cite{Wong:2019kwg},  that confirms the measurements from Supernovae, but the Baryon Acoustic Oscillations (BAO) result agrees with the CMB from Planck ~\cite{Gil-Marin:2016wya}. From these results, the tension even increased to 5.3$\sigma$ by combining all of the $H_0$ values with different origins.

One popular approach to release the $H_0$ tension is to add extra dark radiation or another sterile neutrino in the epoch before recombination. These additional components can make the cosmological expanding faster with a lower sound horizon $r_s$, which lead to a shorter history of the Universe before recombination and a larger $H_0$ value\cite{Wyman:2013lza,Dvorkin:2014lea, Leistedt:2014sia,Ade:2015rim, Lesgourgues:2015wza,Adhikari:2016bei,DiValentino:2016hlg,Canac:2016smv,Feng:2017nss,Oldengott:2017fhy,Lancaster:2017ksf,Kreisch:2019yzn,Li:2020gtk,Zhang:2020mox,Ballardini:2020iws,Novikov:2016fzd, Novikov:2016hrc,Vagnozzi:2019ezj,Alestas:2020zol,Yang:2021eud,DiValentino:2021zxy,Yang:2021flj,Yang:2020zuk,DiValentino:2020zio}.
However, this approach will change the oscillation of acoustic and damping scale~\cite{Hu:1996vq}, which has been constrained by the precise observation of CMB.  So this approach is unable to raise $H_0$ drastically\cite{Aghanim:2018eyx,Raveri:2017jto}.

In another approach, the model named the ``early dark energy (EDE)"  provided a dark component only becomes effective around the recombination time. The EDE model can avoid the previous problem and release the Hubble tension manifest~\cite{Poulin:2018cxd, Braglia:2020iik, Braglia:2020bym, Braglia:2020auw}. However, the EDE model will also change the CMB acoustic peaks and amplitudes by its perturbation.
The Acoustic Dark Energy (ADE) model can avert this trouble of EDE and increase the value of $H_0$\cite{Lin:2019qug, Lin:2020jcb} because of the particular Equation of State (EoS) $w_{ADE}$ within its dark fluid component.


In the ADE model, the sound speed varies with the EoS in the background. This energy density becomes important in the matter radiation equally epoch and impacts the CMB through the gravitational effects of its acoustic oscillations~\cite{ArmendarizPicon:2000ah}.

We explore a new exponential form of EoS for the ADE model in this paper. The new model has the same advantages as ADE but has fewer free parameters called the Exponential Acoustic Dark Energy (eADE) model. 
In this study, we will introduce this eADE model and its fitting results from observational data.
In particular, we use the {\bf CAMB}~\cite{Lewis:1999bs} and {\bf CosmoMC}~\cite{Lewis:2002ah} packages with the Markov chain Monte Carlo (MCMC) method to give the constraints of the model.

This paper is organized as follows.
In Sec.~\ref{sec:model}, we introduce our eADE model, and derive the evolution equations for the dark fluid part in the linear perturbation theory.
In Sec.~\ref{sec:fitting}, we present our numerical calculations. In particular, we show
the CMB power spectra, matter power spectra and constrain the model parameters from several cosmological observation datasets.
Finally, we will present our conclusion in Sec.~\ref{sec:CONCLUSIONS}.

\section{Exponential Acoustic Dark Energy model}
\label{sec:model}

The eADE is defined to be a perfect dark fluid. Its EoS  $w_{eADE}$ and the sound speed $c^2_s$ give the main mark of this model.
We can relieve the Hubble tension by acoustic phenomenology of the linear sound waves from the background and the perturbation in the eADE model.

We define the EoS $w_{eADE}$ to be 
\begin{equation}
\label{eq:eom}
1 + w_{eADE}(a) = 2^{1-\frac{a_c}{2a}},
\end{equation}
where $a_c$ is the moment for critical redshift $z_c$ ($z_c = 1/a_c-1$), which epoach make eADE becomes dominant.
For the value of $w_{eADE} $ equal to $-1$ when $a \ll a_c$,  $w_{eADE} $ equal to $\sqrt{2}-1$ when $a = a_c$, and $w_{eADE} $ equal to  $1$ when $a = 1$. Since the value of $a_c$ is around $\mathcal{O}(10^{-4})$, we regard $a_c$ as $a_c \ll 1$ for the present time.
The development of this EoS is the same as that in ADE model \cite{Lin:2019qug}, which can be seen in the Table~\ref{tab:1}. Moreover, this EoS in our eADE model has one less free parameter than that in the ADE model (parameter $p$ at Eq.1 of ref~\cite{Lin:2019qug}). 
And our model also has a transiently important contribution to the energy density around $a_c$ then decays quickly.
So comparing with the ADE model\cite{Lin:2019qug}, the eADE model not only has a simpler form but also takes a new exponential term in EoS.

In Tab.\ref{tab:1}, we also compared the evolution of the EoS in the EDE model \cite{Poulin:2018cxd} with the parameter $n = 2$ and $n=3$, which shows the different evolutions of EoS in ADE and EDE.

\begin{table}[ht]
	\begin{center}
		\caption{  The benchmark point of EoS for eADE, ADE, and EDE model at different cosmic scale factor $a$. For the EDE model, we select the parameter $n$ equal to 2 and 3. }
		\begin{tabular}{|c|c|c|c|} \hline
			Value of EoS & $a\ll a_c$ &$a= a_c$&$a=1$
			\\ \hline
			$w_{eADE}$& $-1          $&$\sqrt 2 -1$&$1$
			\\ \hline
			$w_{ADE}$& $-1$& $\sqrt 2 -1$&$1$
			\\ \hline
			$w_{EDE} $ with $n=3$& $-1$&$-1/4$&$1/2$
			\\ \hline
			$w_{EDE} $ with $n=2$& $-1$&$-1/3$&$1/3$
			\\ \hline
		\end{tabular}
		\label{tab:1}
	\end{center}
\end{table}

The eADE density evolves as \cite{Poulin:2018dzj},
\begin{equation}
\label{eq:rhode1}
\Omega_{eADE}(a)=\frac{\rho_{eADE}}{\rho_{tot}}= 2 f_c \frac{(c_s^2 +1)^2-(w_{eADE}+1)^2}{(c_s^2+1)^2} 
\end{equation}
where $f_c=\frac{\rho_{eADE}(a_c)}{\rho_{tot}(a_c)}$ means the contribution of eADE at $a_c$.

The pressure of the eADE can be given as $P_{eADE}={{w}_{eADE}}{{\rho}_{eADE}}$.
From Eq.~\ref{eq:eom} and Eq.~\ref{eq:rhode1}, the $P_{eADE}$ can be written naturally as
\begin{equation}
P_{eADE}= \frac{6 H^2 f_c}{8\pi G} \frac{(c_s^2 +1)^2-(w_{eADE}+1)^2}{(c_s^2+1)^2} w_{eADE},
\end{equation}
where the $c_s^2$ is the sound speed which is defined as
\begin{equation}
\label{eq:cs2}
c_s^2 \equiv \frac{\dot{P}_{eADE}} {\dot{\rho}_{eADE}} = \dot w_{eADE}.
\end{equation}
The parametrization of $w_{eADE}$ can get from the cosmological observation. The value of $c_s^2 $ close to 1, and the parameter was fixed by 1 in many previous work\cite{Lin:2019qug, Lin:2020jcb}. So we will also consider $c_s^2$ to be 1 in the following calculation.
In this way, the $\Omega_{eADE}$ and $P_{eADE}$ can be rewritten as, 
\begin{equation}
\label{eq:rhode}
\Omega_{eADE}(a) \overset{c_s^2=1}{=} {f_c} [ 2-\frac{(w_{eADE}(a)+1)^2}{2}].
\end{equation}

Since the eADE model has a main effect in the early Universe around $a_c$, we can consider the $\Omega_{eADE}$ back to zero when $a$ equals $1$ in the present time. 

The pressure term can be shown as,
\begin{equation}
\label{eq:p}
P_{eADE} \overset{c_s^2=1 }{=}\frac{6 H^2 f_c}{8\pi G}(-2\xi^3+\xi^2+2\xi^1-\xi^0),
\end{equation}
where $\xi(a)=2^{-\frac{a_c}{2a}}$. 
It can give a polynomial of the $\xi$, which order in pressure can be shown as a group with 3, 2, 1, and 0 naturally.
And when $a $ is $ 1$, the pressure value of the eADE is equal to zero. It shows the effect of the $P_{eADE}$ is the same as the matter in the late time Universe.

Now we discuss the perturbation part of this model.
The $\dot{\rho}_{eADE}$ can be calculated as
$\delta{\rho}_{eADE}=\delta_{eADE} \rho_{eADE}$, where $\delta_{eADE}$ is the density contrast. Thus, the $\delta P_{eADE}$ can be given by
\begin{equation}
\label{eq:1}
\delta P_{eADE} = (w_{eADE}\delta_{eADE}+\delta w_{eADE})\rho_{eADE}.
\end{equation}

We consider
\begin{equation}
\label{eq:F3}
\delta w_{eADE} = (c_s^2 - w_{eADE})\delta_{eADE}.
\end{equation}
The quantity $\delta w_{eADE}$ is the spatial fluctuations in the EoS, so the pressure fluctuation in Eq.\ref{eq:1} can be rewritten as
\begin{equation}
\label{eq:2}
\delta P_{eADE} =\delta\rho_{eADE} c_s^2.
\end{equation}
The EoS fluctuations can arise from temporal variations in the background EoS through a general coordinate.

We are working on the synchronous gauge and consider the corresponding velocities of the dark fluid as $\theta_{eADE}$.
The conservation equations then become
\begin{eqnarray}
\label{eq:pert1}
&& \dot{\delta}_{eADE}=-(1+w_{eADE})( \theta_{eADE}+\frac{\dot h}{2} )-3\frac{\dot a}{a}( c_s^2-w_{eADE})\delta_{eADE} - k \theta_{eADE}\,, 
\\
\label{eq:pert2}
&& \dot{\theta}_{eADE}= \frac{\dot a}{a}(2w_{eADE}-1)\theta_{eADE}-\frac{\dot w_{eADE}}{1+w_{eADE}}\theta_{eADE}+ \frac{c_s^2 k^2}{a^2(1+w_{eADE})}\delta_{eADE}  \,,
\end{eqnarray}
where $k$ is the $k$-space unit vector, $h$ is the metric perturbation in the Fourier space \cite{Ma:1995ey} and  $\theta_{eADE}=0$ in the rest frame of this dark fluid.

\section{Fitting Result of The Exponential Acoustic Dark Energy}
\label{sec:fitting}

\begin{table}[ht]
	\begin{center}
		\caption{ Priors for cosmological parameters in the exponential acoustic dark energy model.  }
		\begin{tabular}{|c||c|} \hline
			Parameter & Prior
			\\ \hline
			Model parameter $\log(a_c)$& $-10.0 \leq \log(a_c) \leq  -3.0$
			\\ \hline
			Model parameter $f_c$& $0 \leq f_c \leq 1$
			\\ \hline
			Sound speed & $0 \leq c_s^2 \leq 1.5$
			\\ \hline
			Baryon density parameter& $0.5 \leq 100\Omega_bh^2 \leq 10$
			\\ \hline
			CDM density parameter & $0.1 \leq 100\Omega_ch^2 \leq 99$
			\\ \hline
			Optical depth & $0.01 \leq \tau \leq 0.8$
			\\ \hline
			$\frac{\mathrm{Sound \ horizon}}{\mathrm{Angular \ diameter \ distance}}$  & $0.5 \leq 100 \theta_{MC} \leq 10$
			\\ \hline
			Scalar power spectrum amplitude & $2 \leq \ln \left( 10^{10} A_s \right) \leq 4$
			\\ \hline
			Spectral index & $0.8 \leq n_s \leq 1.2$
			\\ \hline
		\end{tabular}
		\label{tab:3}
	\end{center}
\end{table}

\begin{figure}
	\centering
	\includegraphics[ width=0.70 \linewidth]{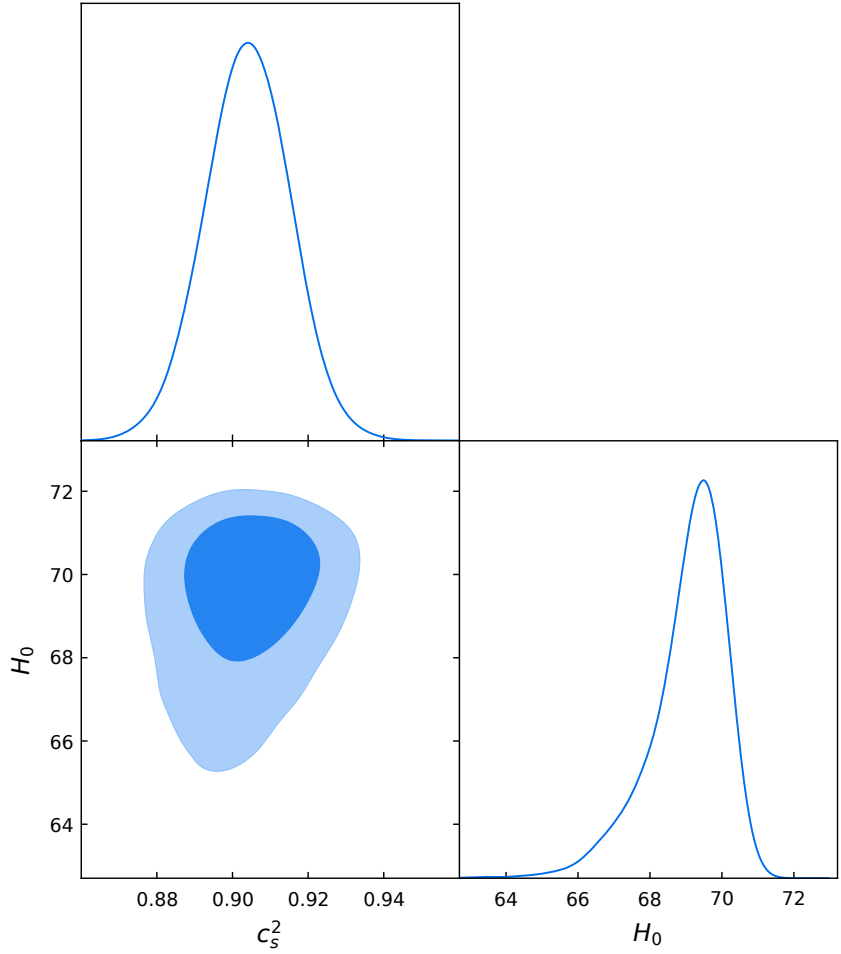}
	\caption{One and two-dimensional distributions of $c_s^2$ and $H_0$ in Planck dataset, where the contour lines represent 68$\%$~ and 95$\%$~ C.L., respectively.}
	\label{fg:cs}
\end{figure}

\begin{figure}
	\centering
	\includegraphics[width=0.90  \linewidth]{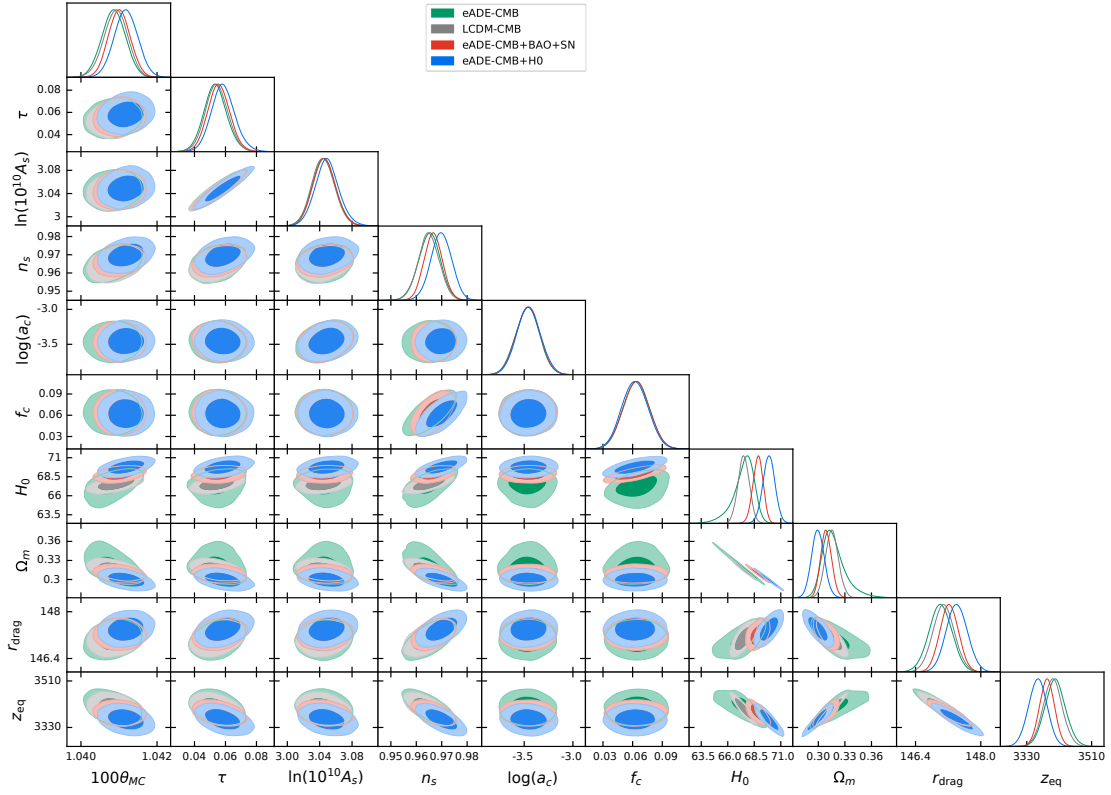}
	\caption{One and two-dimensional distributions of parameters in eADE model and $\Lambda$CDM, where the contour lines represent 68$\%$~ and 95$\%$~ C.L., respectively.}
	\label{fg:h0}
\end{figure}



\begin{table}[ht]
	\begin{center}
		\caption{Fitting results for the eADE model, where the limits are given at 95$\%$ C.L.. We also added the corresponding $\Lambda$CDM result inside brackets. }
		\begin{tabular} {|c|c|c|c|c|}
			\hline
			Parameter & CMB ($c_s^2 \neq 1$) & CMB  & CMB+BAO+SN & CMB+$H_0$ \\
			\hline
			{\boldmath$\Omega_m $} & $0.3176^{+0.0275}_{-0.0238}$& \makecell[c]{$0.3146^{+0.0312}_{-0.0150}$\\$ (0.3153^{+0.0149}_{-0.0140})$} & \makecell[c]{$0.3090^{+0.0125}_{-0.0114}$\\$(0.3104^{+0.0106}_{-0.0103})      $ }& \makecell[c]{$0.2995^{+0.0148}_{-0.0137}$\\ $(0.3077^{+0.0133}_{-0.0091})$}\\
						\hline
			{\boldmath$\log(a_{c})    $} & $-3.31^{+0.30}_{-0.17}        $& $-3.46^{+0.29}_{-0.22}     $&$-3.46^{+0.28}_{-0.22} $&$-3.46^{+0.28}_{-0.22} $\\
						\hline
			{\boldmath$f_c            $} & $0.059^{+0.110}_{-0.058}         $&$0.062^{+0.028}_{-0.031}                $&$0.063^{+0.027}_{-0.030}$&$0.063^{+0.027}_{-0.030}$\\
						\hline
			$c_s^2                       $ & $0.907^{+0.027}_{-0.024}         $& $1$ (fixed) & $1$ (fixed)&  $1$ (fixed)\\
						\hline
			{\boldmath$100 \theta_{MC} $} & $1.04060^{+0.00066}_{-0.00058}$& \makecell[c]{$1.04088\pm {0.00062}$\\$(1.04092^{+0.00059}_{-0.00064})$ }&\makecell[c]{ $1.04093\pm{0.00056} $\\$   (1.04102^{+0.00055}_{-0.00058})$} & \makecell[c]{$1.04108\pm{0.00059}$\\$ (1.04119\pm{0.00060})$}\\
			\hline
			$r_{drag}                       $ & $146.86^{+0.50}_{-0.49}         $& \makecell[c]{$ 146.98^{+0.49}_{-0.50}  $\\$  (147.09^{+0.53}_{-0.51})    $}&\makecell[c]{$ 147.12^{+0.47}_{-0.42}$\\$   (147.23^{+0.46}_{-0.43})  $}&\makecell[c]{ $ 147.41\pm{0.57} $\\$  (147.48\pm{0.49})  $}\\
			\hline
			$H_0                       $ & $69.62^{+1.06}_{-2.09}         $& \makecell[c]{$ 68.21^{+0.95}_{-2.72}  $\\$  (67.36^{+1.05}_{-1.06})    $}&\makecell[c]{$ 68.73^{+0.61}_{-0.65}$\\$   (67.72^{+0.77}_{-0.79})  $}&\makecell[c]{ $ 70.06^{+1.13}_{-1.09} $\\$  (68.58^{+1.10}_{-1.13})  $}\\
			\hline	
			$R-1 $ & $0.01957           $& $0.00797 (0.0103)   $& $0.01969 (0.01667)$& $0.01064 (0.03088)   $\\
			\hline		
			$\chi^2                  $ & $2802.61           $& $2796.53 (2800.70)   $& $3833.05 (3841.84)$& $2814.97 (2817.24)   $\\
			\hline
		\end{tabular}
		\label{tab:cmb+bao+sn}
	\end{center}
\end{table}

\begin{figure}[htbp]
	\centering
    \includegraphics[width=0.6 \linewidth]{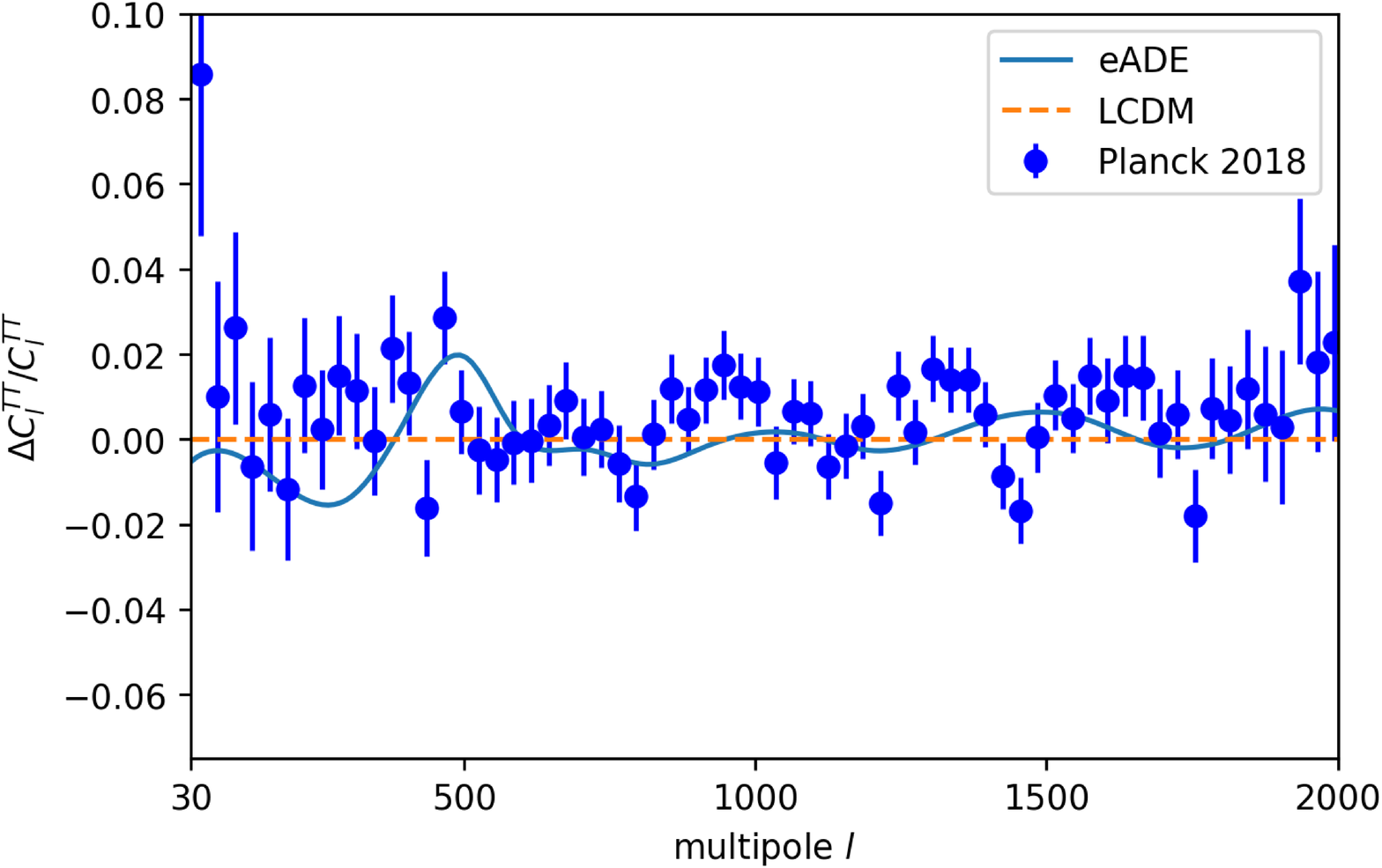}
	\includegraphics[width=0.6 \linewidth]{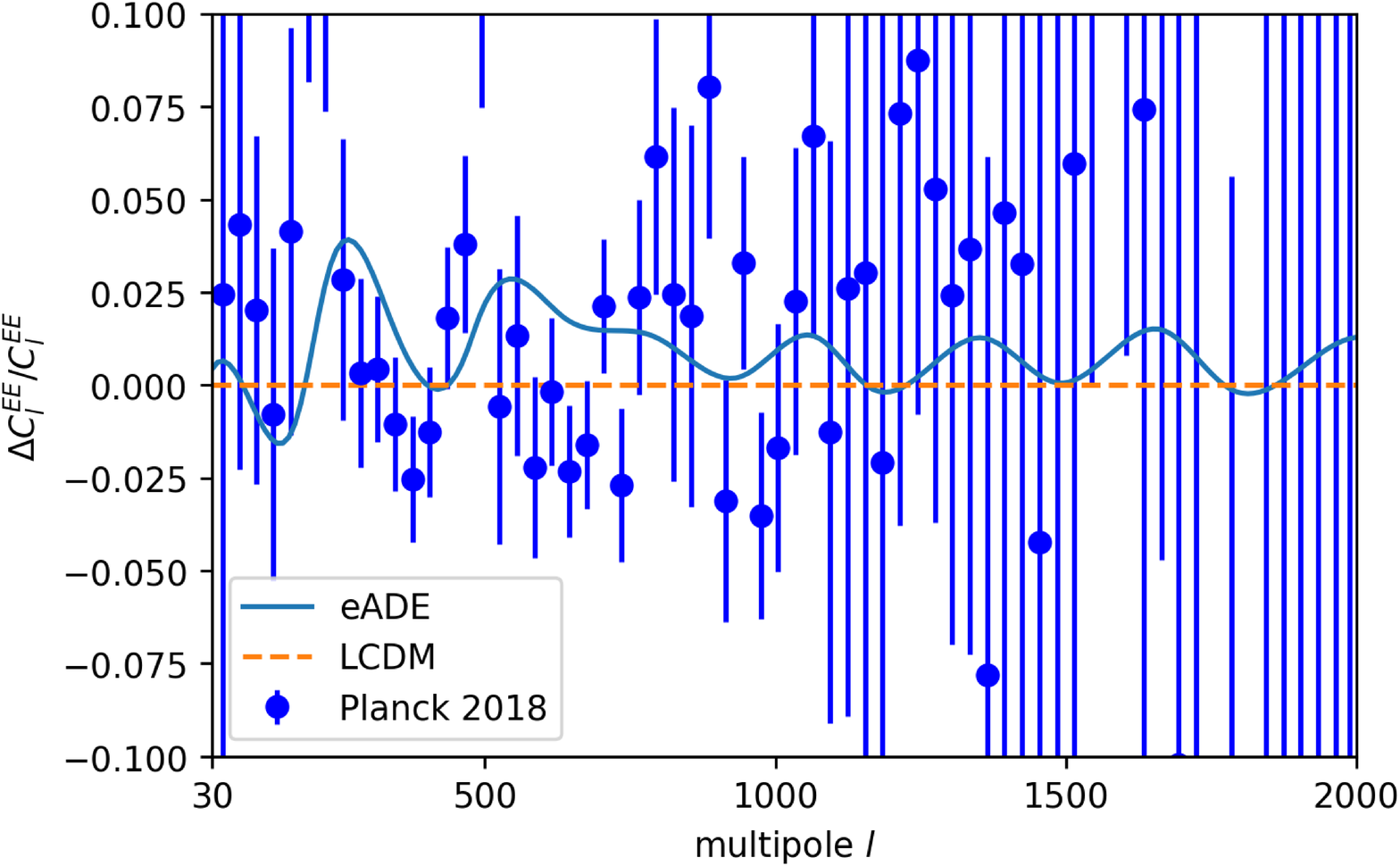}	
	\caption{The $\Delta C_l$ is the change between the eADE model and the $\Lambda$CDM for the CMB power spectra in TT and EE mode.
}
	\label{fg:1}
\end{figure}

\begin{figure}
	\centering
	\includegraphics[width=0.45 \linewidth, angle=270]{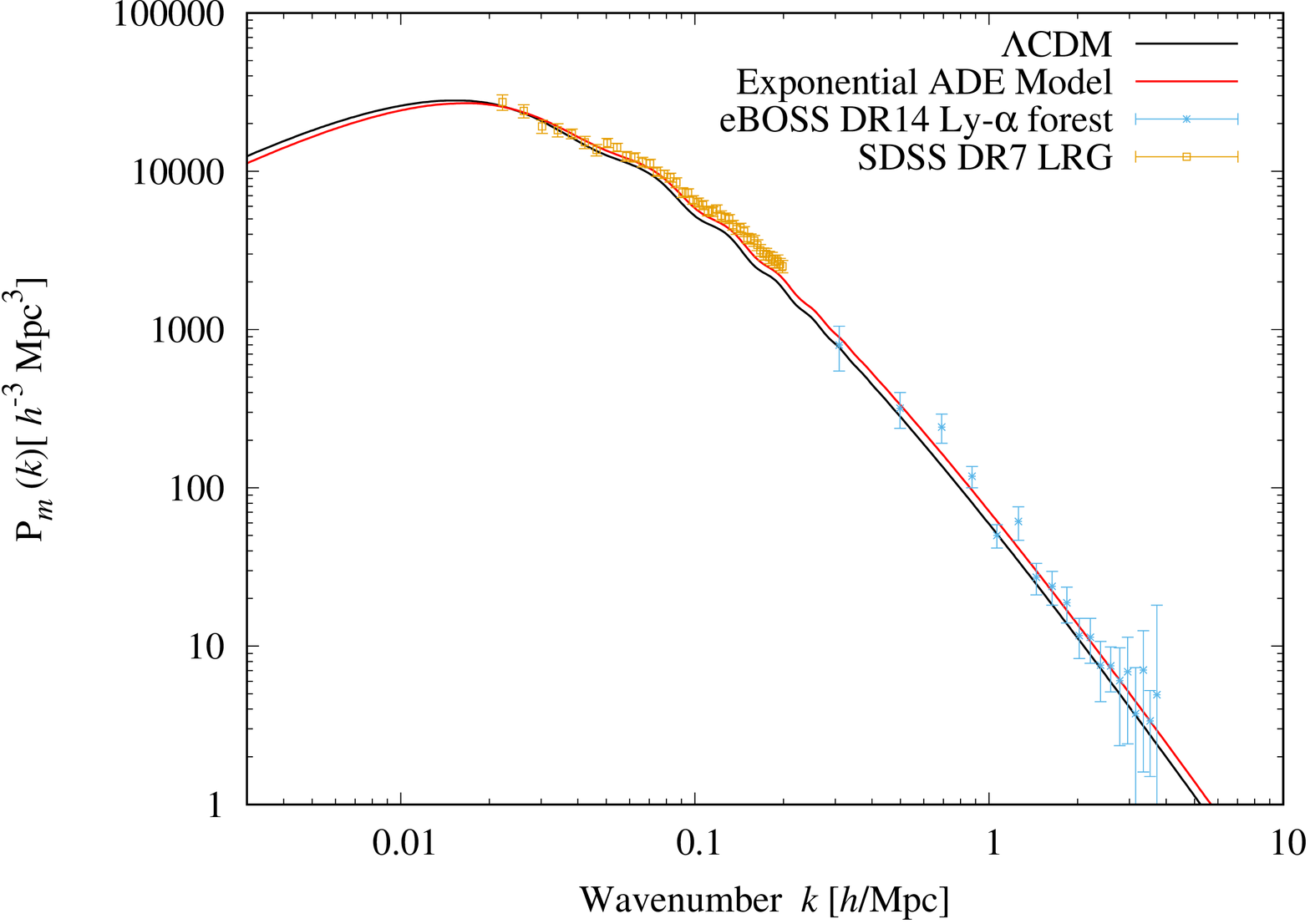}
	\caption{Matter power spectra for the  $\Lambda$CDM and the eADE model with data points in different scales. The Ly-$\alpha$ data from eBOSS clustering on the small scales, and SDSS Luminous Red Galaxy(LRG) clustering on the  intermediate scales. }
	\label{fg:2}
\end{figure}

In this section, we present our Markov Chain Monte Carlo (MCMC) fitting results by  the public code of ${\bf CAMB}$\cite{Lewis:1999bs} and {\bf CosmoMC} program~\cite{Lewis:2002ah}.
The prior of the free parameters were set in Table~\ref{tab:3}.
Note we limit $f_c$ in the range from 0 to 1. And for the parameter $a_c$,
we used the form of $\log(a_c)$ in the global fitting.

For the Cosmic Microwave Mackground (CMB) data, we considered temperature fluctuation from \textit{Planck 2018} with
\texttt{Planck\_highl\_TTTEEE}, \texttt{Planck\_lowl\_EE}, and
\texttt{Planck\_lowl\_TT}, \texttt{Planck\_lensing}
polarization~\cite{Aghanim:2019ame,Akrami:2019izv,Aghanim:2018eyx,Akrami:2018odb,Aghanim:2018oex}.
The baryon acoustic oscillation
(BAO) data in our global fitting included 6dF Galaxy Survey~\cite{Beutler:2011hx} and the Sloan
Digital Sky Survey~\cite{Ross:2014qpa,Alam:2016hwk}.
The Hubble Space Telescope (HST)
result~\cite{Riess:2019cxk} we used is  the single data point of
$H_{0}=74.03 \pm{1.42}$. In addition, the type Ia Supernovae data in the fitting comes from the latest Pantheon result with 1048 data points
\cite{Scolnic:2017caz}. 
Without proper interdependence consideration, the covariance and light curve parameters in SH0ES and Pantheon observation can not valid the constraints simultaneously and independently. 
That leads to a sudden or rapid change in $H(z)$ at $z <0.1$, which would impact both constraints for the Hubble flow and absolute peak magnitude $M_B$.
One method is to use the SNIa distance ladder to calibrate Hubble flow, that constrains and covariance will be contained in the SNIa sample. 
This approach will use in  SH0ES + Pantheon's future data release and cosmological models' constrain~\cite{DiValentino:2021izs}.

First of all, we set the sound speed $c_s^2$ as a free parameter and fit the eADE model with Planck 2018 data. The result of $\Omega_m$, $\log(a_c)$,  $f_c$, $H_0$, and other parameters are shown in Table \ref{tab:cmb+bao+sn}. Especially, we give the contour result of $c_s^2$ and $H_0$ in Fig.\ref{fg:cs}, {which shows the Hubble parameter equals to $69.62^{+1.06}_{-2.09}$ in 95$\%$ C.L.  and can release the $H_0$ tension effectly}. The sound speed has the best-fit value of $c_s^2 = 0.907^{+0.027}_{-0.024}$ and close to 1. So in the next simulation, we set $c_s^2 $ is $1$ and focus on the value of $\log(a_c)$ and $f_c$.

To test the effect of the eADE model in different observations, we consider the combination of the dataset as CMB, CMB+BAO+SN, CMB+$H_0$, separately. 
We present the fitting result of marginalized 1D and 2D posterior distributions in Fig.~\ref{fg:h0} and the limit of 95$\%$C.L. in Tab.\ref{tab:cmb+bao+sn}.


{The gray color of Fig.\ref{fg:h0} is the result of $\Lambda$CDM with Planck dataset, and the green, red and blue color correspond to the eADE in the CMB, CMB+BAO+SN, and CMB+$H_0$ dataset, respectively.} If the  $f_c$ equal to zero, the eADE model will back to $\Lambda$CDM. Lucky, all the $f_c$ and $\log(a_c)$ have two bounds at 95$\%$C.L..
In different data component, the value of $f_c$ still has two bounds with  center value around 0.07 ($c_s^2$ fixed to 1), { that means our model can be  distinguished with the standard cosmological model in 2$\sigma$.  From this figure, we can see the values of parameters $\theta_{MC}$, $r_{drag}$, and $\Omega_m$ in eADE are smaller than that in $\Lambda$CDM, and the $n_s$, $H_0$ and $z_{eq}$ are larger than the corresponding result in $\Lambda$CDM.
}

The eADE model in different data components can enlarge the $H_0$ value from $\Lambda$CDM's result, and this model will release the Hubble tension to a certain extent. For only Planck data, the eADE has $H_0 = 68.21^{+0.95}_{-2.72}$, which larger than the $\Lambda$CDM's result in same dataset ($ 67.36^{+1.05}_{-1.06}  $). In the data of CMB+BAO+SN, the eADE gives the constraint of $ H_0 = 68.73^{+0.61}_{-0.65}$, which is also larger than the result of $ 67.72^{+0.77}_{-0.79}  $ from $\Lambda$CDM. Especially, the eADE model can release the $H_0$ tension noteworthy in the data of Planck+Riess Prior, that $H_0= 70.06^{+1.13}_{-1.09}  $ can distinguish with the corresponding result of $\Lambda$CDM ($ 68.58^{+1.10}_{-1.13}$) in 2$\sigma$. This $H_0$  in eADE is the same as the Freedman's result of Tip of the Red Giant Branch (TRGB)\cite{Freedman:2021ahq, Cerny:2020inj}.


Tab.~\ref{tab:cmb+bao+sn} gives the fitting result of the matter density parameter $\Omega_m$ and  the $\chi^2$ in the $\Lambda$CDM and eADE model. 
In the epoch of $a_c \approx a_{eq}$, the acoustic peak of CMB is impacted by the eADE.
At the background level, because the additional dark energy can enlarge the value of the total energy density before recombination, it will lead to a smaller sound horizon $r_s$ and a shorter expanding history at this time.
In general, the sound horizon can be written as CMB last-scattering as 
\begin{equation}
\label{eq:rs}
r_{\mathrm{s}}=\int_{0}^{t_{\star}} \frac{d t}{a(t)} c_{s}=\int_{z_{\star}}^{\infty} \frac{d z}{H(z)} c_{s},
\end{equation}
where the $t_{\star}$ and $z_{\star}$ correspond to the end of baryon drag epoch, so we has $r_{drag}\approx r_{s}$. In this paper, we can see the difference of sound horizon at the two model in Fig.\ref{fg:h0} and Tab.\ref{tab:cmb+bao+sn}, that shows the $r_{drag}$ in eADE is smaller than the result in $\Lambda$CDM.
In this way, our model has possible to release the $H_0$ tension.
The angular size on the last-scattering surface $\theta_{MC}$ will also have a smaller value in the eADE model. It can give a reduced inverse distance ladder scale, which will lead to the increase of $H_0$. 
This distance measurement calibration modification is not only for CMB observation but also for the BAO and SN through the inverse distance ladder.

Since there are two additional free parameters $a_c$ and $f_c$ in the eADE model, 
we consider the  Akaike Information Criterion (AIC)~\cite{Akaike1974} 
to statistical analysis and comparison of the two models.
The AIC estimator can be
expressed by
\begin{equation}
AIC \equiv -2 \ln L_{max}+2k 
\end{equation}
where $L_{max}\equiv p(d|\theta_{max}, M)$ is the maximum likelihood
value, and $k$ is the number of free parameters.  
Since $\chi^2$ equal to $-2 \ln L_{max}$, the $\chi^2$ value of the eADE model in CMB+BAO+SN dataset is 3833.05,  which is smaller than the result in $\Lambda$CDM (3841.84).
The difference of $\chi^2$ between the two models is 8.79, it means
we can get
$\triangle AIC= \triangle AIC_{\Lambda CDM}-\triangle AIC_{eADE}=
4.79$ in the CMB+BAO+SN dataset. Using the same method, we can obtain the $\triangle AIC=0.17$ in only CMB observation and $\triangle AIC=-1.73$ from CMB+$H_0$ dataset.
The value of $\triangle AIC$ shows a preference of the eADE model in CMB+BAO+SN compared with the $\Lambda$CDM  model~\cite{Arevalo:2016epc}.


In Fig.\ref{fg:1}, we can see the difference of CMB power spectra between the eADE model and $\Lambda$CDM in TT mode and EE mode. The data points of CMB in blue color come from Planck 2018\cite{Aghanim:2019ame}, { the theoretical value of $\Lambda$CDM (orange dash line) and eADE (grey blue line) comes from fitting results in Planck data. 
In the TT mode, the eADE result is closer to the Planck data points in the first peak,  which shows the advantages of the eADE model.
Meanwhile, the EE mode resulting from Planck still has too much noise in the high $l$, so the eADE model has a smaller error with some data points than the standard model but is hard to distinguish with the $\Lambda$CDM.}

In Fig.~\ref{fg:2}, we can compare the eADE and $\Lambda$CDM in matter power spectra with different observations. The Luminous Red Galaxy (LRG) data come from SDSS DR7~\cite{Reid:2009xm}, which is in the redshift between 0.6 and 1.0. The data points are closer to the eADE model prediction than the  $\Lambda$CDM especially at high wavenumber $k$ [$h/Mpc$]. On the small scale, we compared the two models with the eBOSS DR14 Ly-$\alpha$ forest data \cite{Abolfathi:2017vfu}. The center value of the Ly-$\alpha$ data points also prefers eADE in high $k$ [$h/Mpc$]. Since the error bar of the $Ly-\alpha$ forest dataset are still too large, we can not distinguish the eADE model and the $\Lambda$CDM in matter power spectra.

\section{Conclusions}
\label{sec:CONCLUSIONS}

In this work, we have introduced a new additional perfect dark fluid, named the eADE model, as a candidate to release the $H_0$ tension.
The eADE model was defined with a special EoS as an exponential function of the sound speed. This model is different from $\Lambda$CDM at the matter radiation equally time.

We calculated the background evolution and the linear perturbation of the eADE model and have performed the MCMC global fitting with the latest observational data which includes Planck, SDSS, HST, and Pantheon.
From this simulation, the eADE model can release the $H_0$ tension with the result of 
$H_0 = 70.06^{+1.13}_{-1.09}$  which is contrasted with the $H_0 = 68.58^{+1.10}_{-1.13} $ in $\Lambda$CDM at  95$\%$~ C.L. in Planck+$H_0$.
The fitting result of eADE shows a smaller $\chi^2$ than that in $\Lambda$CDM. The  AIC analysis with different datasets also confirms this result.
Finally, we compared the eADE model with $\Lambda$CDM in the CMB and the matter power spectra. The observational data points prefer our eADE model to $\Lambda$CDM at a small scale structure but are hard to distinguish in present precision.

\section*{Acknowledgments}

The author thanks Prof. G. Tasinato for the inspirer.
We also thank Prof. Xin Zhang and Dr. Shu-Lei Ni for the useful discussion.

This research was supported by Basic Science Research Program through the National Research Foundation of Korea(NRF) funded by the Ministry of Education through the Center for Quantum Spacetime (CQUeST) of Sogang University (NRF-2020R1A6A1A03047877).

\end{document}